\documentclass[preprint,3p, review]{elsarticle} 
\usepackage{multicol,caption}

\newenvironment{Figure}
 {\par\medskip\noindent\minipage{\linewidth}}
 {\endminipage\par\medskip}

\usepackage{amssymb}

\usepackage{lineno}
\usepackage{graphicx}
\usepackage{dcolumn}
\usepackage{bm}
\usepackage{rotating}
\usepackage{float}
\usepackage{amsmath}
\usepackage{amssymb}
\usepackage{color}
\usepackage{multirow}
\usepackage{booktabs}

\usepackage{subfig}
\usepackage{graphicx}

\usepackage{ccfonts}
\usepackage{times}
\usepackage[utf8]{inputenc}
\biboptions{sort&compress}
\usepackage[]{siunitx}

\journal{Physics Letters B}

\begin{document}

\begin{frontmatter}

\title{Measurement of proton electromagnetic form factors in the time-like region using initial state radiation at BESIII}

\begin{small}

\author{
M.~Ablikim$^{1}$, M.~N.~Achasov$^{10,c}$, P.~Adlarson$^{67}$, S. ~Ahmed$^{15}$, M.~Albrecht$^{4}$, R.~Aliberti$^{28}$, A.~Amoroso$^{66A,66C}$, M.~R.~An$^{32}$, Q.~An$^{63,49}$, X.~H.~Bai$^{57}$
% ...
\\
\vspace{0.2cm}
(BESIII Collaboration)\\
\vspace{0.2cm} {\it
$^{1}$ Institute of High Energy Physics, Beijing 100049, People's Republic of China\\
% ...
$^{4}$ Bochum Ruhr-University, D-44780 Bochum, Germany\\
% ...
$^{10}$ G.I. Budker Institute of Nuclear Physics SB RAS (BINP), Novosibirsk 630090, Russia\\
% ...
$^{15}$ Helmholtz Institute Mainz, Staudinger Weg 18, D-55099 Mainz, Germany\\
% ...
$^{28}$ Johannes Gutenberg University of Mainz, Johann-Joachim-Becher-Weg 45, D-55099 Mainz, Germany\\
% ...
$^{32}$ Liaoning Normal University, Dalian 116029, People's Republic of China\\
% ...
$^{49}$ State Key Laboratory of Particle Detection and Electronics, Beijing 100049, Hefei 230026, People's Republic of China\\
% ...
$^{57}$ University of Jinan, Jinan 250022, People's Republic of China\\
% ...
$^{63}$ University of Science and Technology of China, Hefei 230026, People's Republic of China\\
% ...
$^{66}$ University of Turin and INFN, (A)University of Turin, I-10125, Turin, Italy; (B)University of Eastern Piedmont, I-15121, Alessandria, Italy; (C)INFN, I-10125, Turin, Italy\\
$^{67}$ Uppsala University, Box 516, SE-75120 Uppsala, Sweden\\
% ...
$^{c}$ Also at the Novosibirsk State University, Novosibirsk, 630090, Russia\\
% ...
}
}

\end{small}
%
\begin{abstract}
The electromagnetic process $e^{+}e^{-}\to p\bar{p}$ is studied with 
the initial-state-radiation technique using 7.5 fb$^{-1}$ of data
collected by the BESIII 
experiment at seven energy points 
from 3.773 to 4.600~GeV. The Born cross section and the effective 
form factor of the proton are measured 
from the production threshold to 3.0~GeV/$c^{2}$
using the $p\bar{p}$ invariant-mass spectrum.
The ratio of electric 
and magnetic form factors of the proton is determined from the analysis 
of the proton-helicity angular distribution.
\end{abstract}
\end{frontmatter}

\begin{multicols}{2}

\section{Introduction}
The investigation of nucleon structure through electromagnetic probes plays 
a central role in the understanding of strong interactions. 
Space-like (SL) photons (momentum transfer squared $q^2  <$  0) 
in elastic electron-nucleon scattering experiments allow an accurate description 
of the three-dimensional structure of the nucleon through the study of the electromagnetic 
form factors (FFs).
The electric FF $G_{\rm E}$ and the magnetic FF $G_{\rm M}$ are assumed to be analytic functions of $q^2$~\cite{drell1961}, and thus are 
also defined for the time-like (TL) kinematic domain $q^2 > 0$. 
Furthermore, it is possible to relate SL and TL FFs through dispersion 
relations~\cite{belushkin}. In the TL region, nucleon FFs can be associated 
with the time evolution of the charge and magnetic distributions inside the nucleon~\cite{kuraev}.

Compared to the SL sector, where a percent level precision has been achieved~\cite{punjabi}, 
the knowledge of the TL proton FF is rather limited. 
Proton FFs in the TL region have been studied by various experiments in the direct 
annihilation processes $e^{+}e^{-}\to p\bar{p}$  
~\cite{adone73, dm182, dm283, dm290, fenice93, fenice94, fenice98, cleo05, bes205, bes3rscan, cmd3, Ablikim:2019eau}  
and $ p\bar{p} \to e^{+}e^{-} $ ~\cite{e760, Ba94, E835,Andreotti:2003bt}, 
and in the initial-state-radiation (ISR) process 
$e^{+}e^{-}\to p\bar{p}\gamma$~\cite{babar06, babar13t, babar13u, Ablikim:2019njl}. 
Due to low statistics, many previous experiments have only determined 
the absolute value of the effective FF of the proton from the cross-section measurement. 
More recent measurements~\cite{bes3rscan, cmd3, Ba94, babar06, babar13t} have been able 
to determine the ratio of the proton FF absolute values ($R_{ \rm em}=|G_{\rm E}|/|G_{\rm M}|$) 
in the  $p\bar{p}$ invariant mass ($M_{p\bar{p}}$) region below 3.08~GeV/$c^{2}$.  
The best determination of $R_{\rm em}$, with a precision of around $10\%$, 
has been achieved by BESIII~\cite{Ablikim:2019eau}.

The ISR technique with an undetected photon has been used in our previous 
study~\cite{Ablikim:2019njl} of the process $e^{+}e^{-}\to p\bar{p} \gamma$ 
to measure the TL proton FFs. In that analysis, events were selected where the ISR photon 
was emitted at small polar angles (SA-ISR), and hence the threshold 
region below 2 GeV/$c$ was not accessible due to the limited 
angular acceptance of the BESIII tracking system. In this Letter we extend our previous study 
to the case where the ISR photon is emitted at large polar angles (LA-ISR) 
and is  detected. This allows access to the threshold region 
and provides measurements of the  proton helicity angle $\theta_{p}$
in the full $M_{p\bar{p}}$ range, 
in contrast to the analysis of the SA-ISR events. By analyzing the distribution 
of $\theta_{p}$, defined in the process $e^{+}e^{-}\to p\bar{p} \gamma$ 
as the angle between the proton momentum in the $p\bar{p}$ rest frame 
and the momentum of the $p\bar{p}$ system in the $e^{+}e^{-}$ c.m.\ frame, 
it is possible to determine the ratio of the proton FFs.
\begin{Figure}
\centering
\includegraphics[scale=0.45]{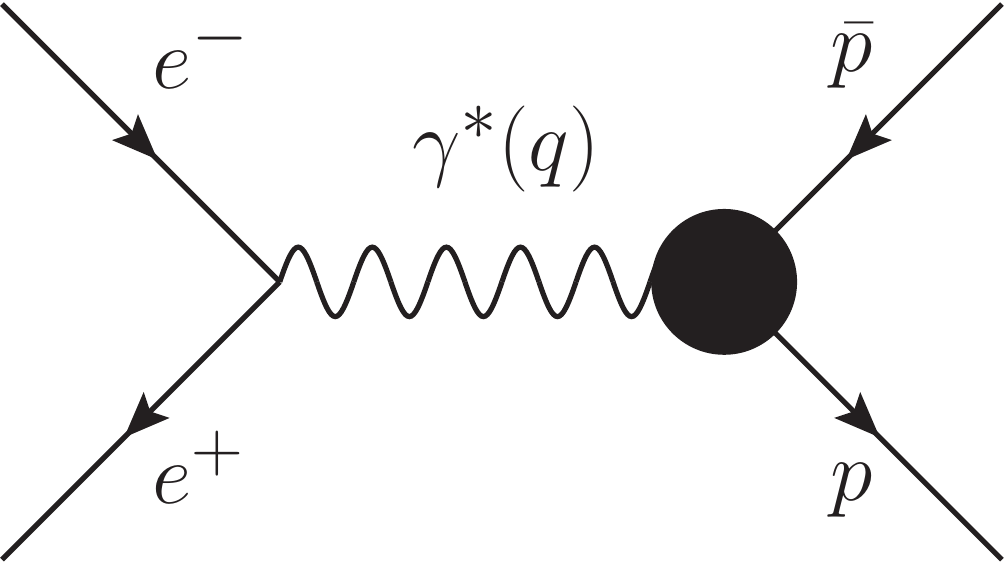}
\captionof{figure}{Feynman diagram for the process $e^+ e^- \to p\bar{p}$ under the assumption of one virtual photon exchange. The bold vertex provides access to the hadronic FFs information.}
\label{fig:dxs}
\end{Figure}

In the $e^{+}e^{-}$ center-of-mass (c.m.) frame, the differential cross section for the process $e^{+}e^{-}\to p\bar{p}$ under the assumption of one virtual photon exchange (Figure~\ref{fig:dxs}) is~\cite{zichichi}:
\begin{linenomath*}
\begin{equation}
 \begin{aligned}
  \frac{d\sigma_{p\bar{p}}(q^{2})}{d\cos\vartheta} = \frac{\pi\alpha^{2}\beta\mathcal{C}}{2q^2}[|G_{\rm M}(q^2)|^{2}(1+\cos^{2}\vartheta) \\ + \frac{|G_{\rm E}(q^2)|^{2}}{\tau}\sin^{2}\vartheta],
 \end{aligned}
 \label{eq:dxs}
\end{equation}
\end{linenomath*}
where $q^{2}$ is equal to the square of the ${p\bar{p}}$ invariant mass $M_{p\bar{p}}$, $\alpha\approx\frac{1}{137}$ is the fine structure constant, $\beta=\sqrt{1-1/\tau}$ is the velocity of the proton with $\tau=q^{2}/4m^{2}_{p}$ and $m_{p}$ the proton mass, and $\vartheta$ is the polar angle of the proton in the $e^{+}e^{-}$ c.m. frame  where the z-axis points along the direction of the positron momentum. The Coulomb factor $\mathcal{C}=\frac{y}{1-e^{-y}}$ with $y=\frac{\pi\alpha}{\beta}$ accounts for the electromagnetic interaction between the outgoing proton and antiproton~\cite{rinaldo12, babar13t}. The cross section depends on the moduli of the magnetic and electric FFs, which can be determined from the analysis of the proton angular distribution. The precise knowledge of the FFs in a wide kinematic region probes the transition region, from non-perturbative to perturbative QCD (pQCD).

By integrating the differential cross section (Eq.~(\ref{eq:dxs})), the total cross section for the process $e^{+}e^{-}\to p\bar{p}$ is obtained,
\begin{linenomath*}
\begin{equation}
 \sigma_{p\bar{p}}(q^{2}) = \frac{4\pi\alpha^{2}\beta\mathcal{C}}{3q^2}[|G_{\rm M}|^{2} + \frac{|G_{\rm E}|^{2}}{2\tau}].
 \label{eq:txs}
\end{equation}
\end{linenomath*}
An effective FF is introduced as a linear combination of $|G_{\rm M}|^{2}$ and $|G_{\rm E}|^{2}$,
\begin{linenomath*}
\begin{equation}
 \begin{aligned}
  |G_{\rm eff}| = \sqrt{\frac{2\tau|G_{\rm M}|^{2}+|G_{\rm E}|^{2}}{2\tau+1}},
 \end{aligned}
 \label{eq:geff}
\end{equation}
\end{linenomath*}
which is equivalent to $|G_{\rm M}|$ determined under the assumption of $|G_{\rm M}|=|G_{\rm E}|$.
A complementary approach to study the $e^{+}e^{-}\to p\bar{p}$ process is provided by the ISR technique. This technique makes use of the emission of at least one high energy photon off the beam particles (Figure~\ref{fig:isrxs}) which reduces the invariant mass of the $p\bar{p}$ system in the final state. The differential cross section for the ISR process is
\begin{linenomath*}
\begin{equation}
 \begin{aligned}
  \frac{d\sigma_{p\bar{p}\gamma}(q^{2})}{dq^{2}} &= \frac{1}{s}W(s, x)\sigma_{p\bar{p}}(q^{2}), \\
  W(s, x) &= \frac{\alpha}{\pi x}(\ln\frac{s}{m_{e}^{2}}-1)(2-2x+x^{2}), \\
  x&=\frac{2E^{*}_{\gamma}}{\sqrt{s}} = 1 - \frac{q^{2}}{s},
 \end{aligned}
 \label{eq:isrxs}
\end{equation}
\end{linenomath*}
where $E^{*}_{\gamma}$ is the energy of the ISR photon in the $e^{+}e^{-}$ c.m.\ frame, $W(s, x)$~\cite{bon71} is the radiator function which gives the probability of ISR photon emission, and $m_{e}$ and $\sqrt{s}$ are the electron mass and the c.m.\ energy of the beams, respectively. In the study of the $e^+ e^- \to p\bar{p}\gamma$ process, the cross section for the process $e^{+}e^{-}\to p\bar{p}$ and the ratio of the proton FFs can be measured over the full $M_{p\bar{p}}$ range from the $p\bar{p}$ threshold to $\sqrt{s}$. 
\begin{Figure}
\centering
\includegraphics[scale=0.45]{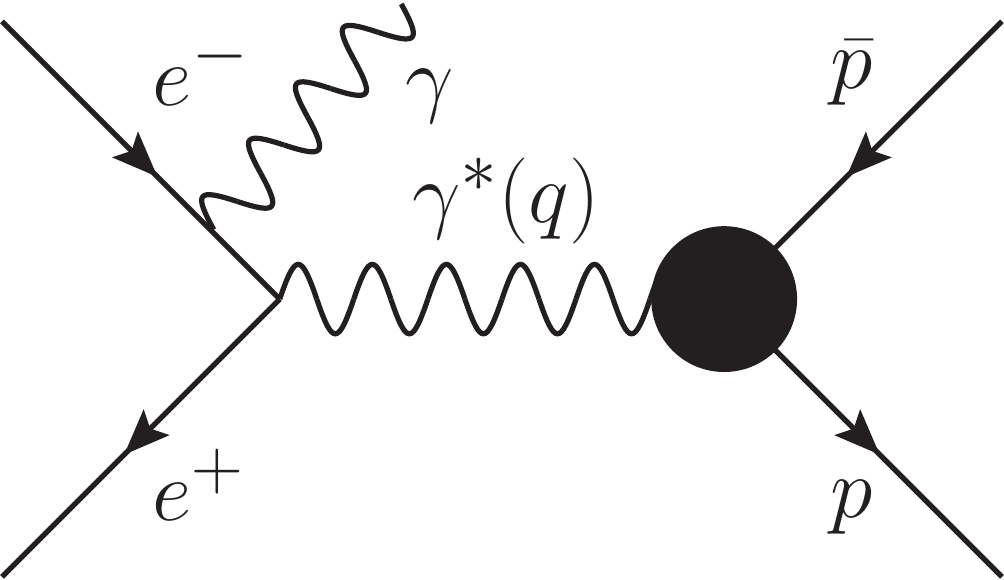}
\captionof{figure}{Lowest order Feynman diagram of the process $e^+ e^- \to p\bar{p}\gamma$ assuming one virtual photon exchange, where $\gamma$ is a real photon emitted from the initial state.}
\label{fig:isrxs}
\end{Figure}
%
\section{BESIII experiment and data sets}
The BESIII experiment collects data at the BEPCII electron-positron collider, which operates at 
 c.m.\ energies between $\sqrt{s}$ = 2.0 and 4.7~GeV.
The baryon TL FFs can be measured at BESIII both in ISR 
and in direct annihilation processes~\cite{bes2020}.
In this letter the investigation of the ISR process $e^{+}e^{-}\to p\bar{p}\gamma$ is reported. 
The data sets used in this analysis \cite{ps2pLumi, xyzLumi} have been collected 
by BESIII at seven c.m.\ energies between 3.773 and 4.600~GeV with a total integrated luminosity of 7.5~fb$^{-1}$ (see Table~\ref{tab:isrData}).

The cylindrical BESIII detector~\cite{bes3det} covers 93\% of the 4$\pi$ solid angle 
around the interaction point (IP), where the electron and positron beams collide 
at a small angle of 22 mrad. Beginning with the innermost component, BESIII consists of the following subdetectors:
(1) a helium-based Main Drift Chamber (MDC) 
composed of 43 cylindrical layers coaxial with the beam pipe, 
(2) a Time-of-Flight system (TOF) consisting of 176 plastic scintillator counters 
in the barrel part, and 96 counters in the endcaps, 
(3) an Electro-Magnetic Calorimeter (EMC) consisting of a barrel and two endcaps 
with 6240 CsI(Tl) crystals, and (4) a Muon Counter (MUC) composed of nine Resistive 
Plate Chamber (RPC) layers in the barrel and eight RPC layers in each endcap. The MDC provides a 
momentum resolution of 0.5\% for charged tracks with 1~GeV/$c$ momentum 
and a spatial resolution of \SI{135}{\micro\meter}. 
The resolution of the energy-loss measurement (d$E$/d$x$) by the MDC is better than 6\%. 
The time resolution of the TOF is 80~ps in the barrel and 110~ps in the endcaps. 
The EMC provides an energy measurement with a resolution of 2.5\% in the barrel and 5\% 
in the endcaps for photons/electrons with an energy of 1~GeV. 
The MUC is used to identify muons and provides a spatial resolution better than 2~cm. 
The MUC is housed in the return yoke of the solenoidal magnet, which provides 1~T magnetic field.

Monte Carlo (MC) signal and background samples, simulated using the {\sc Geant4}-based~\cite{geant4a,geant4b}  {\sc BESIII Object Oriented Simulation Tool} (BOOST) software~\cite{bossref}, are used to optimize the event selection criteria, estimate the background contamination and determine the selection efficiency. The signal process $e^{+}e^{-}\to p\bar{p}\gamma$ is generated with the event generator PHOKHARA 9.1~\cite{phokhara}, which includes radiative corrections of ISR up to next-to-leading order, final-state-radiation  and vacuum polarization. Inclusive MC samples generated with BesEvtGen event generator~\cite{conexc01} are used to simulate all the hadronic final states containing $u$, $d$ and $s$ quarks. The dominant background channel, $e^{+}e^{-}\to p\bar{p}\pi^{0}$, is generated exclusively using the phase space  {\sc Conexc}~\cite{conexc01} generator.
\begin{table}[H]
\centering
 \captionof{table}{The integrated luminosity ${\cal L}$ of the data sets used in the $e^{+}e^{-}\to p\bar{p}\gamma$ 
  analysis. The uncertainties are statistical and systematic, respectively.}
 \begin{tabular*}{\columnwidth}{c @{\extracolsep{\fill}} r}
  \toprule
  \hline
  $\sqrt{s}$ [GeV]  & \multicolumn{1}{c}{$\mathcal{L}$ [pb$^{-1}$]} \\
  \hline
  3.773 & 2931.8 $\pm$ 0.2  $\pm$ 13.8~\cite{ps2pLumi}\\
  4.009 & 481.96 $\pm$ 0.01  $\pm$  4.68~\cite{xyzLumi}\\
  4.230 & 1053.9 $\pm$ 0.1   $\pm$ 7.0~\cite{xyzLumi}\\
  4.260 & 825.67 $\pm$  0.13  $\pm$  8.01~\cite{xyzLumi}\\
  4.360 & 539.84  $\pm$ 0.10  $\pm$ 5.24~\cite{xyzLumi}\\
  4.420 & 1041.3 $\pm$ 0.1  $\pm$ 6.9~\cite{xyzLumi}\\
  4.600 & 585.4  $\pm$ 0.1  $\pm$ 3.9~\cite{xyzLumi}\\
  \hline
  \bottomrule
\end{tabular*}
 \label{tab:isrData}
\end{table}
%
\section{Selection of $e^{+}e^{-}\to p\bar{p}\gamma$ events}
\label{eventselection}
Two charged tracks with net charge zero are required in the MDC. The point of closest approach to the IP for each of the tracks is required to lie within a \SI{1}{\centi\meter} radius in the plane perpendicular to the beam and $\pm$\SI{10}{\centi\meter} along the beam direction. The polar angle of the track with respect to the direction of the positron beam, $\theta$, must be inside in the fiducial volume of the MDC, $|\cos\theta|<0.93$. The particle identification (PID) from the relevant sub-detectors is combined to calculate probabilities for the pion, kaon, proton, electron and muon hypotheses for the tracks. The two charged tracks must be identified as a proton and an antiproton. 
In addition, the ratio $E/p$, with the energy $E$ measured in the EMC and the momentum $p$ measured in the MDC, is required to be smaller than 0.5 for the proton candidate to suppress the ISR Bhabha background. Photon candidates are selected using the information 
on the electromagnetic showers in the EMC. It is required that the shower time is within 700 ns of the event start time to suppress electronic noise and energy deposits unrelated to the event. A photon candidate is selected  if its deposited energy is greater than 25 MeV (50 MeV) in the barrel (endcap) region. The barrel (endcap) region is defined as $|\cos\theta^{*}_{\gamma}|<0.80$ ($0.86<|\cos\theta^{*}_{\gamma}|<0.92$), where $\theta^{*}_{\gamma}$ is the photon polar angle. At least one high energy photon is required in the EMC with deposited energy higher than 0.4~GeV. The highest energy photon is then assumed to be the ISR photon candidate.

After the event reconstruction, a four-constraint (4C) kinematic fit is performed 
requiring the four-momentum conservation between the initial $e^{+}e^{-}$ system
and the final $p\bar{p}\gamma$ system.
Events are selected as $e^{+}e^{-}\to p\bar{p}\gamma$ candidates 
if they fulfill the requirement $\chi^{2}_{\rm 4C}<50$. 
The background from the process $e^{+}e^{-}\to p\bar{p}\pi^{0}$ 
can not be completely removed by means of the kinematic fitting, 
and a dedicated background evaluation is performed, as described in the next section.

\begin{figure}[H]
\begin{center}
  \includegraphics[scale=0.43]{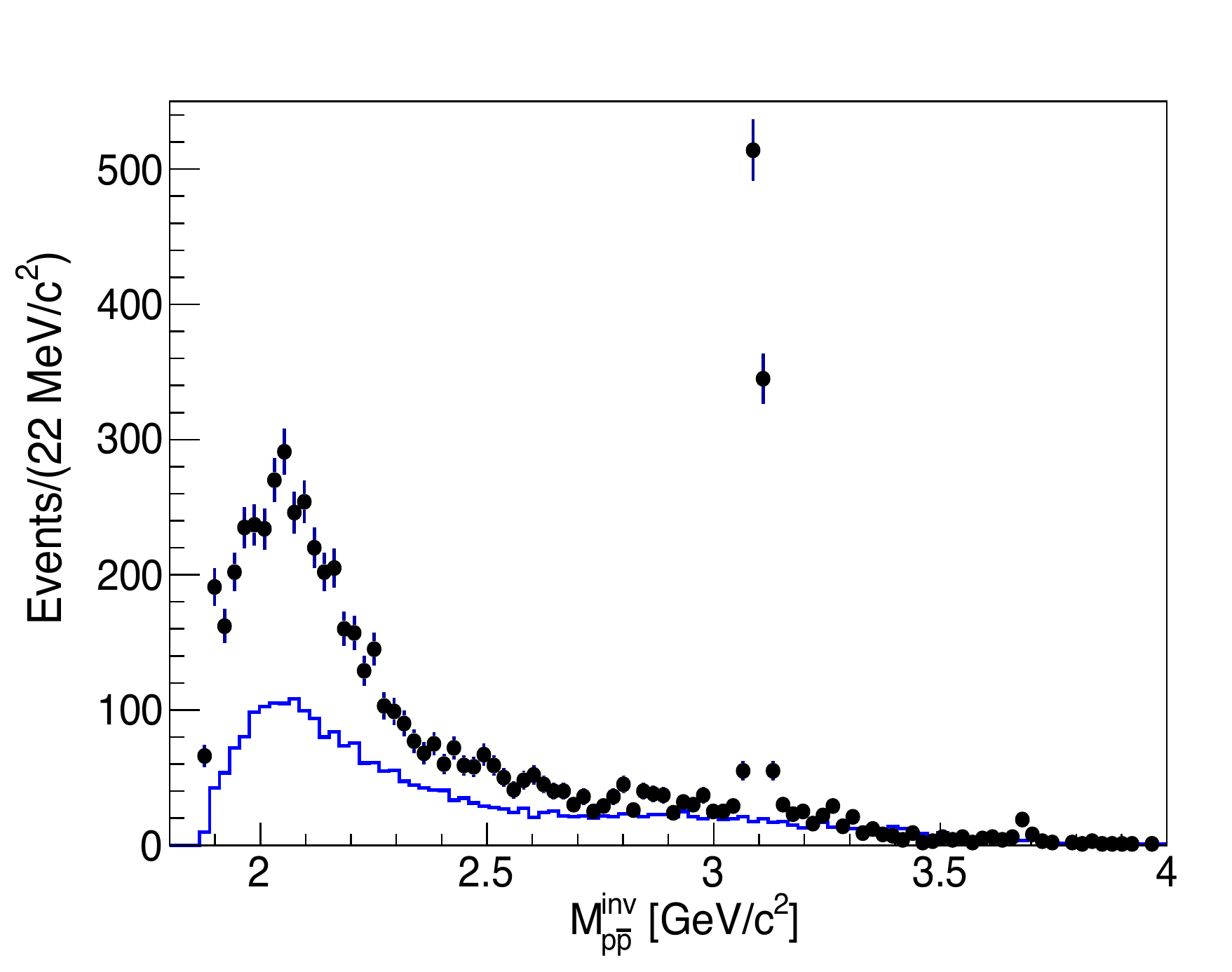}
 \caption{The $M_{p\bar{p}}$ spectrum for the full data sets (black dots with error bars), and the remaining $e^{+}e^{-}\to p\bar{p}\pi^{0}$ background (blue histogram).}
 \label{fig:invMass}
\end{center}
\end{figure}

Figure~\ref{fig:invMass} shows the combined $M_{p\bar{p}}$ spectrum for $p\bar{p}\gamma$ candidates selected at the seven energy points. The residual $e^{+}e^{-}\to p\bar{p}\pi^{0}$ background discussed in Section~\ref{bkg} is also shown (blue histogram in Figure~\ref{fig:invMass}). 
A clear peak from the resonance decay of $J/\psi\to p\bar{p}$ is seen in the spectrum.
By fitting the $J/\psi$ peak using a Breit-Wigner function convolved with a Gaussian, the number of  resonance  decays $ J/\psi\to p\bar{p} ~ (N_{J/\psi})$ for each data sample is determined. The branching fraction $J/\psi \to p \bar p $ can be  calculated  for each data sample individually as follows~\cite{Benayoun:1999hm}:
\begin{linenomath*}
\begin{equation}
\Gamma_{e^+e^-} \times {\cal B}(J/\psi \to p \bar p)=\frac{s m_{J/\psi}}{12 \pi^2}\frac{N_{J/\psi}}{ \epsilon_{J/\psi} W(s,x_{J/\psi}){\cal {L}}},
\end{equation}
\end{linenomath*}
where $m_{J/\psi}$ is the mass of the resonance,  $W(s,x_{J/\psi})$ is the radiator function ($x_{J/\psi}=1-M_{J/\psi}^2/s$), $\Gamma_{e^+e^-}$  is the electronic width of  the ${J/\psi}$~\cite{pdg2016} and ${\cal {L}}$ is the integrated luminosity collected at the given c.m. energy $\sqrt{s}$.  The detection efficiency  $\epsilon_{J/\psi}$ is determined from the signal MC simulations in the $M_{p\bar{p}}$ interval around the $J/\psi$ resonance.
The average value of ${\cal B}({J/\psi \to p \bar p  })$  obtained is  (2.13$\pm$0.09)$\times10^{-3}$, where the uncertainty is statistical only. This result is in a good agreement with the PDG value, (2.12$\pm$0.03)$\times10^{-3}$~\cite{pdg2016}.
%
\section{Background estimation}
\label{bkg}
For the processes 
$e^+ e^- \to \pi^{+}\pi^{-}\gamma$, $K^{+}K^{-}\gamma$, 
$e^{+}e^{-}\gamma$ and $\mu^{+}\mu^{-}\gamma$ no MC events survive
the selection cuts described in Section~\ref{eventselection}. 
The residual background from these sources can be neglected, given the number
of generated MC event exceeds the number of events expected in data.

The main source of background for the process under study is $e^{+}e^{-}\to p\bar{p}\pi^{0}$. 
To ensure a good description of this background in the simulation, 
the MC distributions of $M_{p\bar{p}}$ and $\cos\theta_{p}$ for the 
background process $e^{+}e^{-}\to p\bar{p}\pi^{0}$ are corrected 
by selecting data samples containing only the high-purity $p\bar{p}\pi^{0}$ final state. 
The $p\bar{p}\pi^{0}$ event selection uses the same criteria as the signal selection 
for the charged tracks and requires at least two photons in the EMC without 
the minimum energy cut of 0.4 GeV. After imposing the selection criteria for the charged tracks and photons, a  five-constraint (5C) kinematic fit is performed 
for the  $e^{+}e^{-}\to p\bar{p}\gamma \gamma$ hypothesis 
with requirements of energy and momentum conservation and the invariant mass 
of the two photon candidates being constrained to the $\pi^0$ mass~\cite{pdg2016}. 
The value of $\chi^{2}_{\rm 5C}$ is required to be less than 60. To reject $p\bar{p}\gamma$ events, 
a 4C kinematic fit is performed to the proton, antiproton and the highest energy photon. 
The value of $\chi^{2}_{\rm 4C}$ is required to be larger than 25. 
This selection provides 
a clean $p\bar{p}\pi^{0}$ event sample with the
$e^{+}e^{-}\to p\bar{p}\gamma$ contamination
below 0.1\% level, as estimated from the MC simulation. 
To calculate the remaining $p\bar{p}\pi^{0}$ background 
contamination in the selected $p\bar{p}\gamma$ signal candidates, a weighting method is applied:
\begin{linenomath*}
\begin{equation}
 \mathcal{N}^{\rm bkg} = \mathcal{N}_{\pi^{0}}^{\rm dat} \times \frac{\mathcal{N}_{\rm isr}^{\rm MC}}{\mathcal{N}_{\pi^{0}}^{\rm MC}},
 \label{eq:bckgWeight}
\end{equation}
\end{linenomath*}
where $\mathcal{N}^{\rm bkg}$ is the estimated number of remaining $p\bar{p}\pi^{0}$ 
background events, $\mathcal{N}_{\pi^{0}}^{\rm dat}$ 
and $\mathcal{N}_{\pi^{0}}^{\rm MC}$ are the numbers 
of $p\bar{p}\pi^{0}$ events selected from data and $p\bar{p}\pi^{0}$ MC samples, respectively, and $\mathcal{N}_{\rm isr}^{\rm MC}$ is the number of events selected as $p\bar{p}\gamma$ from the $p\bar{p}\pi^{0}$ MC samples. 

The  background estimate $\mathcal{N}^{\rm bkg}$ is calculated 
in two-dimensional intervals 
of $\cos\theta_{p}$ and $M_{p\bar{p}}$ distributions 
used for the determination of $R_{\rm em}$. For the determination 
of the Born cross section for the process  $e^{+}e^{-}\to p\bar{p}$, $\mathcal{N}^{\rm bkg}$ is calculated
in intervals of $M_{p\bar{p}}$. 
The distribution of  $\mathcal{N}^{\rm bkg}$ as a function  of $M_{p\bar{p}}$, summed over the seven c.m. energy points, is shown as the blue histogram in Figure~\ref{fig:invMass}.
%
%
%
\section{Ratio of proton form factors}
\label{ratioem}
The ratio of the electric and magnetic FFs $R_{\rm em}$ is determined 
by analyzing the distribution of $\cos\theta_{p}$ 
\begin{linenomath*}
\begin{equation}
 \begin{aligned}
 \frac{dN}{d\cos\theta_{p}} = \mathcal{A}[\mathcal{F}_{\rm M}(\cos\theta_{p}, M_{p\bar{p}}) + \\ \frac{R_{\rm em}^{2}}{2\tau}\mathcal{F}_{\rm E}(\cos\theta_{p}, M_{p\bar{p}})].
 \end{aligned}
 \label{eq:pAngDist}
\end{equation}
\end{linenomath*}
where $N$ is the number of selected $e^{+}e^{-}\to p\bar{p}\gamma$ candidates after  $e^{+}e^{-}\to p\bar{p}\pi^{0}$ background subtraction. The shapes of the magnetic contribution 
$\mathcal{F}_{\rm M}(\cos\theta_{p}, M_{p\bar{p}})$ 
and the electric contribution 
$\mathcal{F}_{\rm E}(\cos\theta_{p}, M_{p\bar{p}})$ 
are determined from the MC simulation, 
which includes the radiative corrections.
The distributions obtained for $\mathcal{F}_{\rm M}$ and $\mathcal{F}_{\rm E}$ 
in a given $M_{p\bar{p}}$ interval are approximately proportional 
to $1+\cos^{2}\theta_{p}$ and $\sin^{2}\theta_{p}$, respectively, 
as follows from Eq.~(\ref{eq:dxs}).  The factor $\frac{1}{2}$ arises 
from the normalization of $\mathcal{F}_{\rm M}$ and $\mathcal{F}_{\rm E}$ 
to the same integral, and the parameter $\mathcal{A}$ 
is an overall normalization factor. 
In the fit function, $\tau$ is calculated 
as the mean value over the $p\bar{p}$ mass interval.

\begin{figure}[H]
\centering
 \centering
 \includegraphics[scale=0.19] {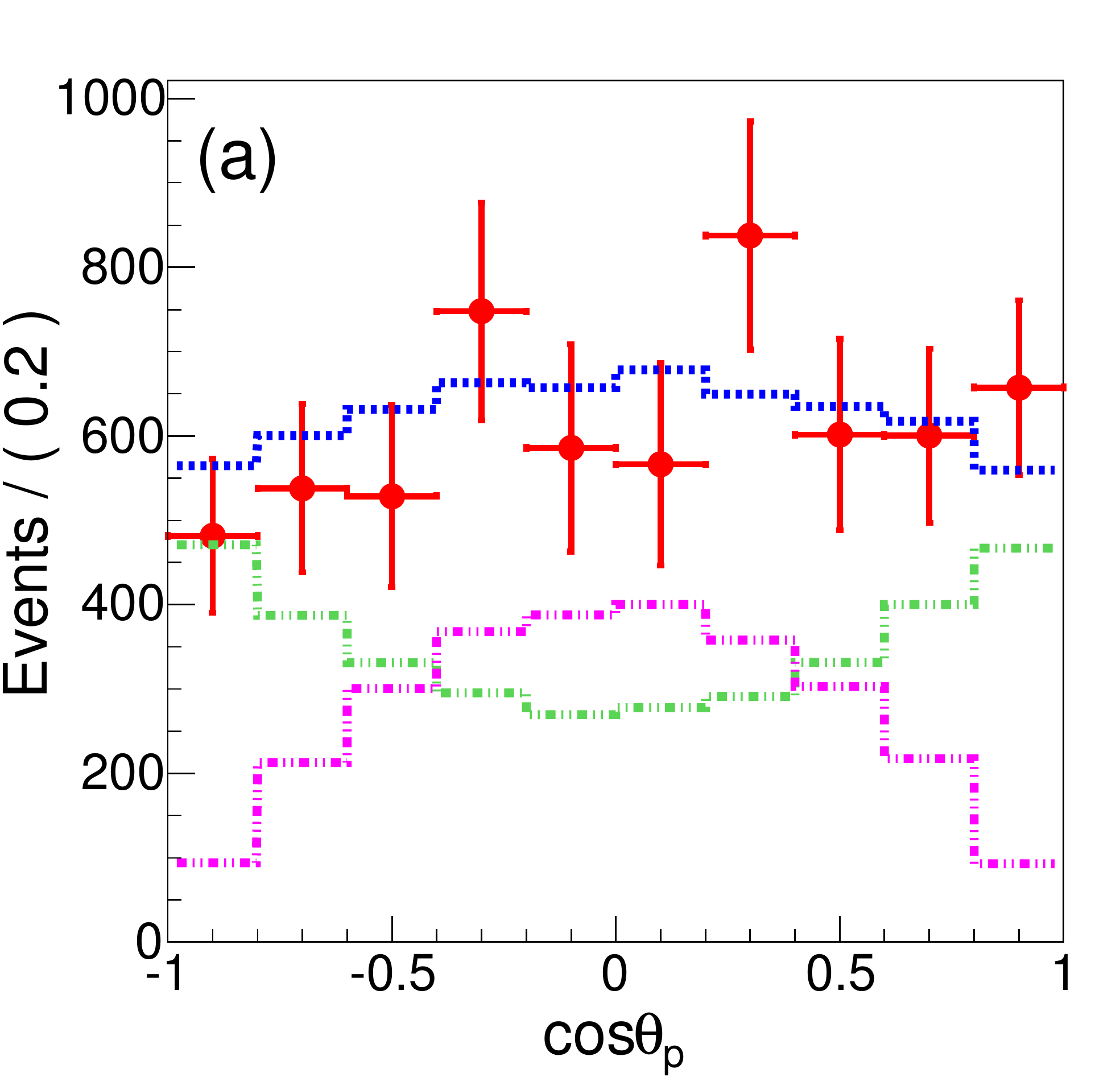}
 \includegraphics[scale=0.19] {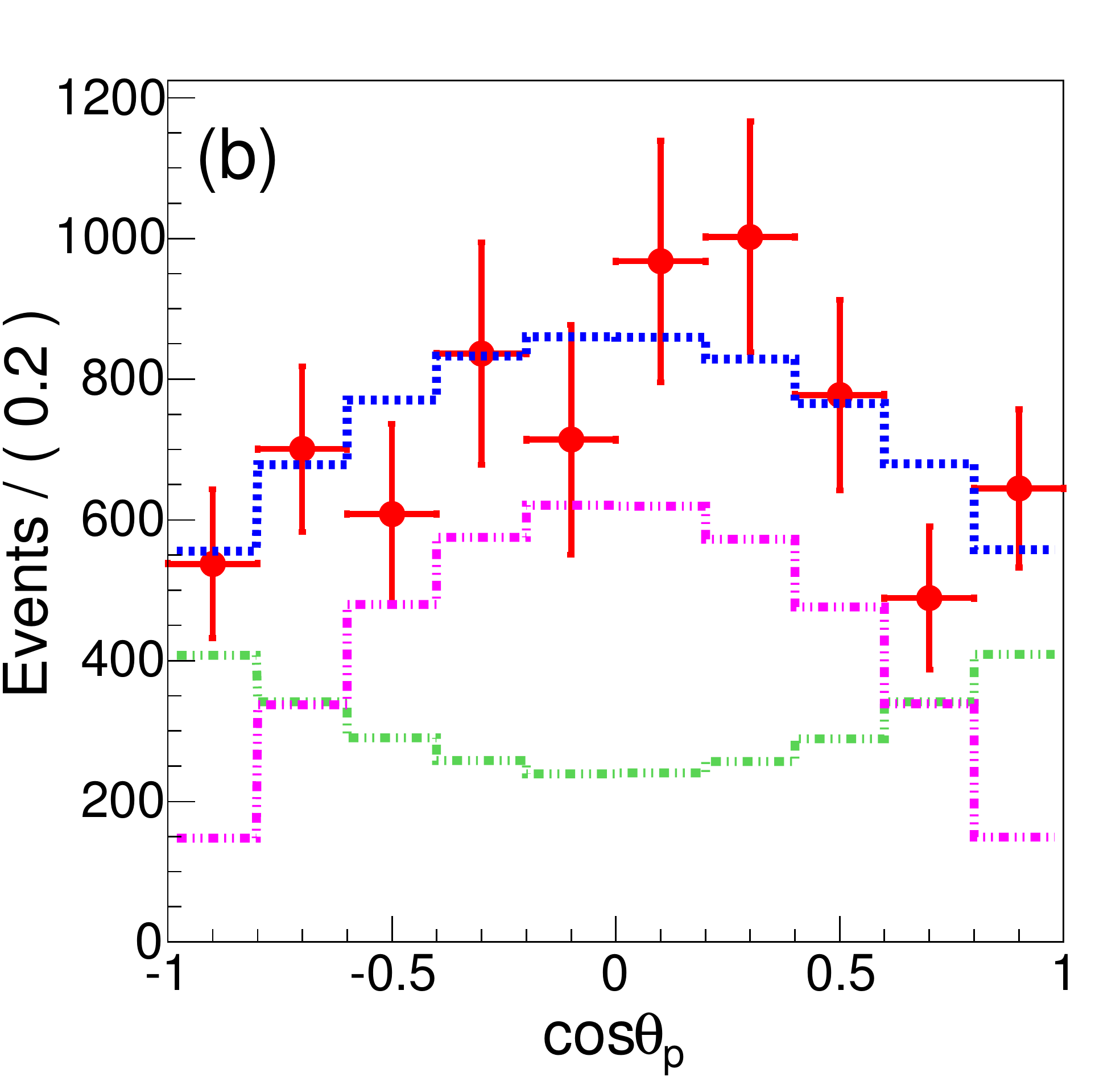} \\
 \includegraphics[scale=0.19] {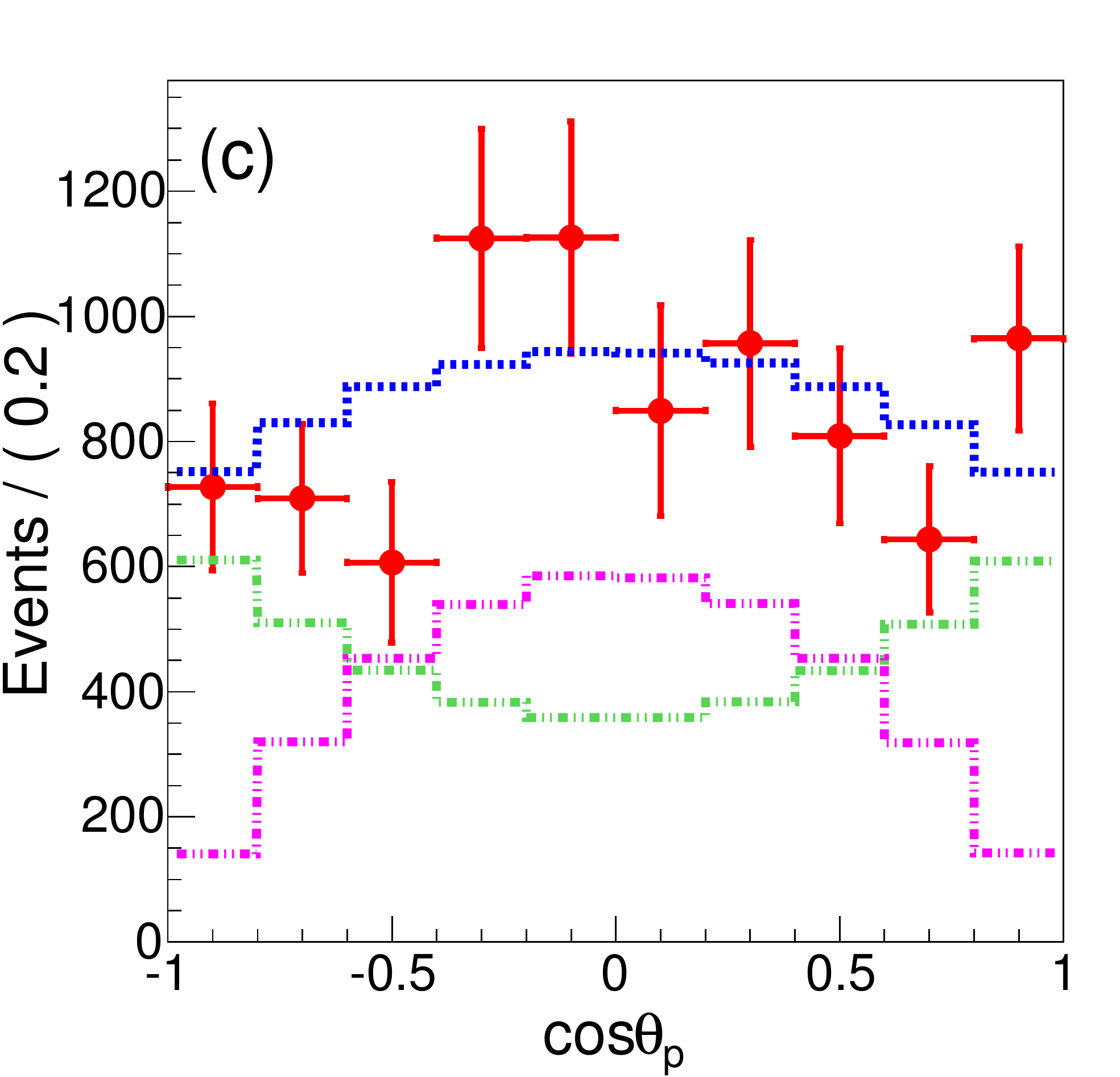}
 \includegraphics[scale=0.19] {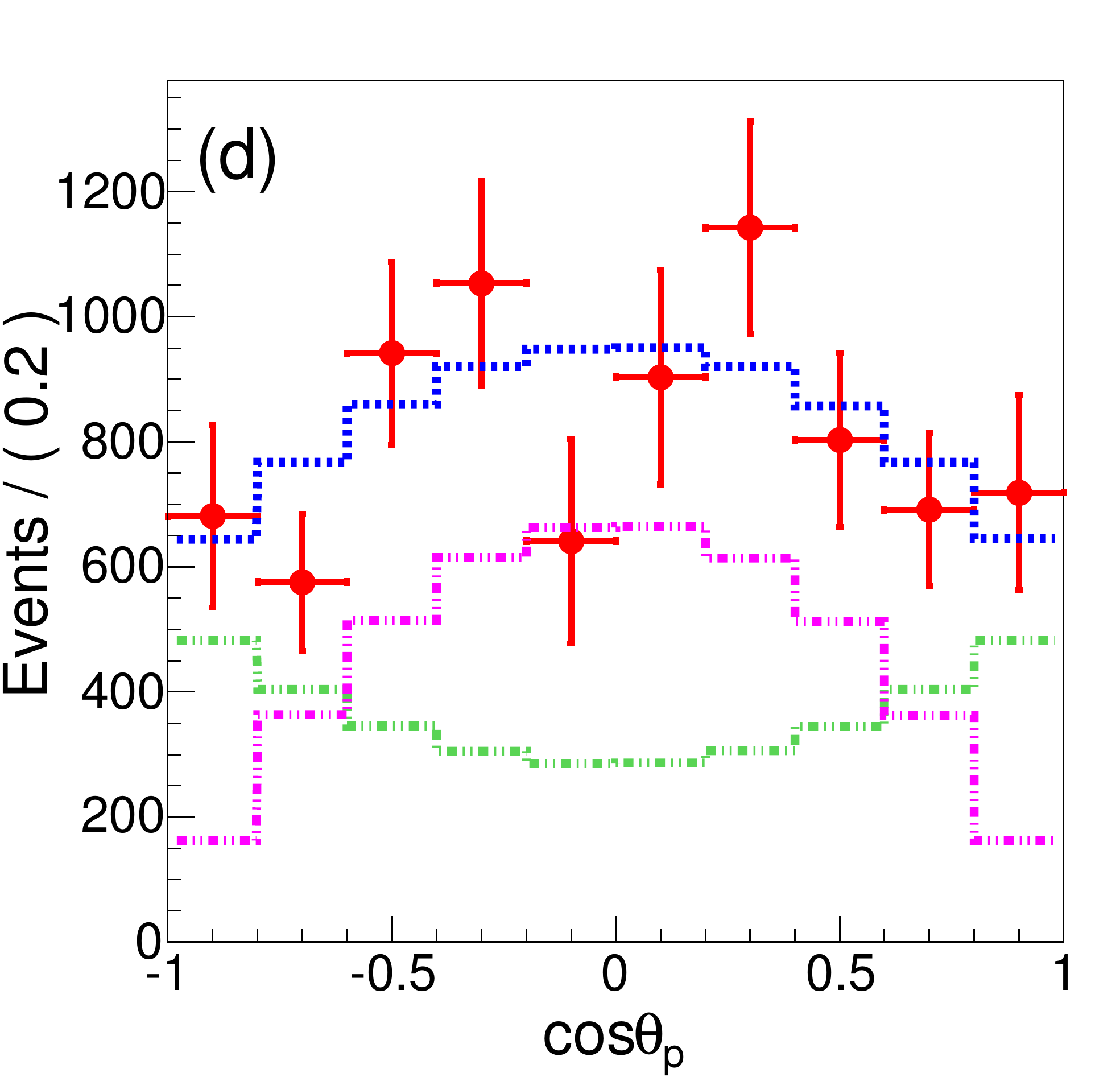} \\
 \includegraphics[scale=0.19] {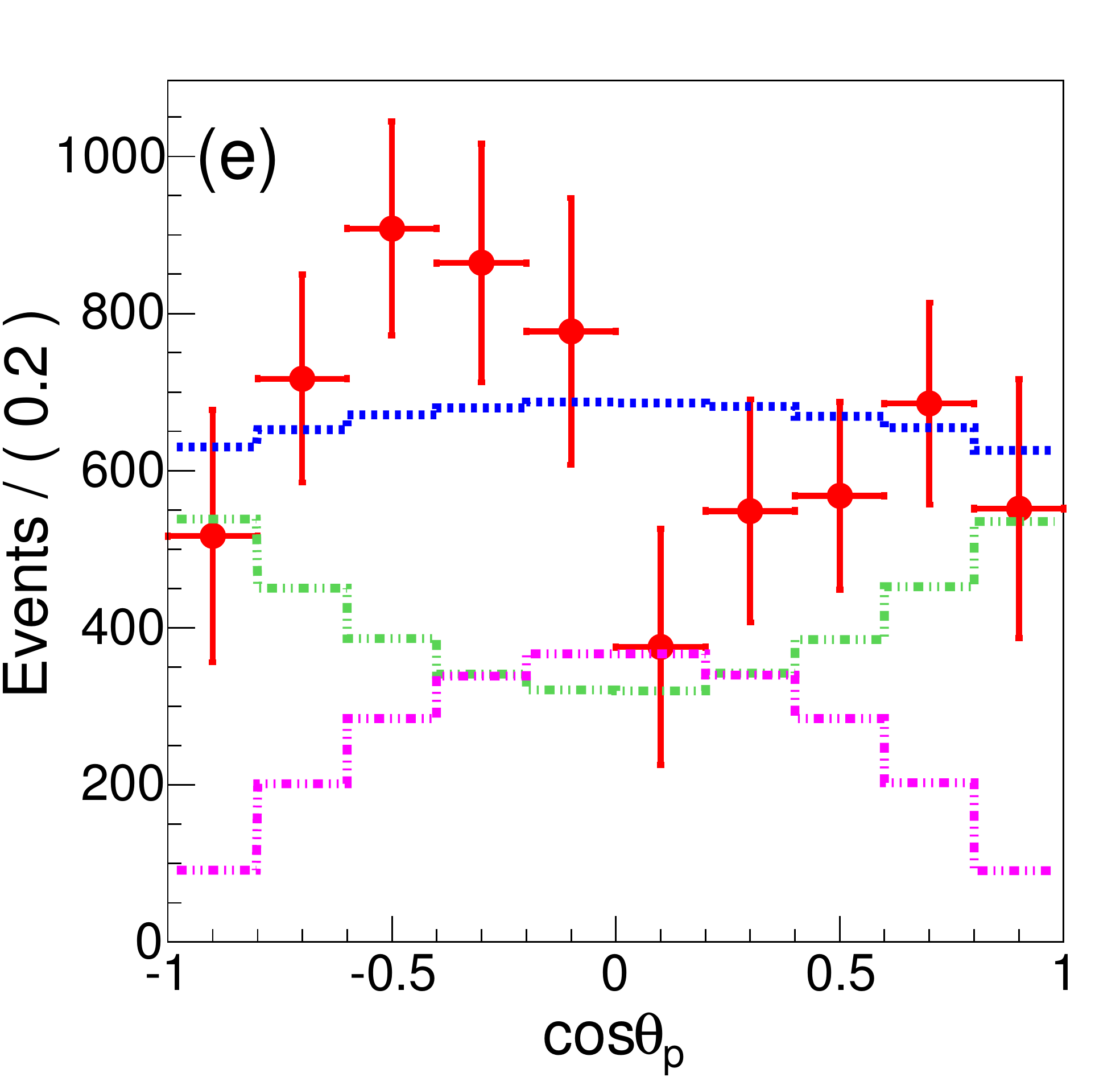}
 \includegraphics[scale=0.19] {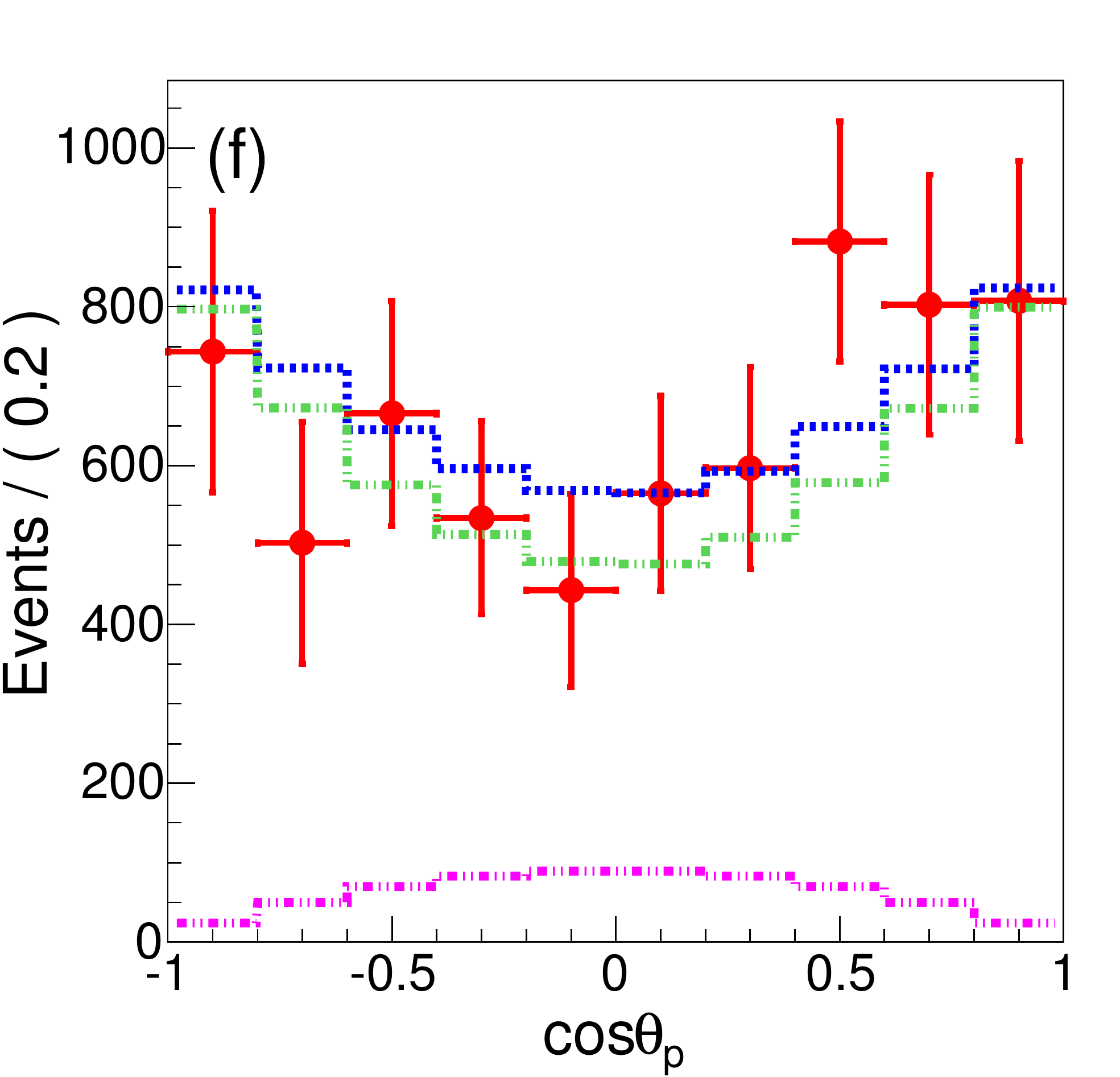}
 \caption{The efficiency-corrected $\cos\theta_{p}$ distribution for different $M_{p\bar{p}}$ intervals after combining all data sets, (a) threshold--1.95~GeV/$c^{2}$, (b) 1.95--2.025~GeV/$c^{2}$, (c) 2.025--2.10~GeV/$c^{2}$, (d) 2.10--2.20~GeV/$c^{2}$, (e) 2.20--2.40~GeV/$c^{2}$ and (f) 2.40--3.00~GeV/$c^{2}$. The red dots represent the distribution from data after $e^{+}e^{-}\to p\bar{p}\pi^{0}$ background subtraction (as described in the text) and MC efficiency correction, the blue dotted lines show the fit to the data, and the green and purple dash-dotted histograms represent the magnetic and electric contributions to the fit, respectively.}
 \label{fig:ratioFit}
\end{figure}

The ratio of the proton FFs is determined 
by fitting the $\cos\theta_{p}$ distribution
in six $M_{p\bar{p}}$ intervals 
from threshold to 3.0~GeV/$c^{2}$,
with $M_{p\bar{p}}$ reconstructed 
from the measured tracks of the proton and antiproton candidates.
For each data set and $M_{p\bar{p}}$ interval, 
the estimated background from 
the process $e^{+}e^{-}\to p\bar{p}\pi^{0}$ is subtracted from 
the number of signal candidates. The remaining signal 
is corrected for the selection efficiency
calculated with corresponding MC samples
and the efficiency-corrected distributions
from all data sets are combined. 
Figure~\ref{fig:ratioFit} 
shows the $\cos\theta_{p}$ distribution for six $M_{p\bar{p}}$ 
intervals, and the results of the fits 
using Eq.~(\ref{eq:pAngDist}). Table~\ref{tab:ratioRes} summarizes 
the $R_{\rm em}$ ratios 
obtained from the fits.
\begin{table}[H]
\centering
\footnotesize
 \captionof{table}{Sum  of the numbers of selected $e^+ e^- \to p\bar{p}\gamma$ 
candidates ($\mathcal{N}_{k}$) and the number of $e^{+}e^{-}\to p\bar{p}\pi^{0}$ 
background events ($\mathcal{N}^{\rm bkg}_{k}$) over the seven c.m. energy points, 
and the results for the ratio $R_{\rm em}$ obtained from the fit for each $M_{p\bar{p}}$ interval $k$.}
 \begin{tabular*}{0.475\textwidth}{c @{\extracolsep{\fill}} ccc}
  \toprule
  \hline
    Mass Int.      & \multirow{2}{*}{$\mathcal{N}_{k}$}  &  \multirow{2}{*}{$\mathcal{N}_{k}^{\rm bkg}$}  &  $R_{\rm em}$  \\

  [GeV/$c^{2}$]   &                                 &                                            &  ($|G_{\rm E}|/|G_{\rm M}|$)  \\
  \midrule
1.877-1.950  &  584  &  160$\pm$5  & 1.27$\pm$ 0.23 $\pm$ 0.09\\ 
1.950-2.025  &  802  &  314$\pm$7  & 1.78$\pm$ 0.33$\pm$0.11 \\
2.025-2.100  &  912  &  352$\pm$8  & 1.46$\pm$ 0.27 $\pm$ 0.09\\
2.100-2.200  &  905  &  370$\pm$8  & 1.82$\pm$ 0.37$\pm$ 0.24\\
2.200-2.400  &  929  &  477$\pm$9  & 1.36$\pm$ 0.37 $\pm$0.22 \\
2.400-3.000  &  1159 & 656$\pm$10  & 0.64$\pm$ 0.47 $\pm$ 0.22\\
  \hline
  \bottomrule
\end{tabular*}
 \label{tab:ratioRes}
\end{table}

The total systematic uncertainties 
for the measurement of the ratio of the proton FFs 
are listed in Table~\ref{tab:ratioRes}. The uncertainties 
include contributions from three main sources: 
the number of $\cos\theta_{p}$ intervals in the fit, 
the 4C-kinematic fit and 
the uncertainty of the  $e^{+}e^{-}\to p\bar{p}\pi^{0}$ 
background estimation. 
The contribution from the number of $\cos\theta_{p}$ intervals
is evaluated by using $\cos\theta_{p}$ distributions
with eight intervals instead of ten. The variations of the fit results
are taken as systematic uncertainties. 
The systematic uncertainty from the 4C-kinematic fit is caused
by inconsistencies between $\chi^{2}_{4C}$ distributions 
in data and simulation. It is estimated by modifying the helix 
parameters of the charged tracks for the MC samples 
according to the method described in Ref.~\cite{pullm}. 
Clean samples of selected
$e^{+} e^{-} \to p\bar{p}\pi^{0}$ events are used to 
compare the $\chi^{2}_{4C}$ distributions for data and MC simulation. 
The difference in the determined $R_{\rm em}$ with and without this 
modification is taken as the systematic uncertainty. 
To determine the uncertainty from the background estimation, 
the number of background events
obtained in Section~\ref{bkg} is varied by one standard deviation
and the fit results after subtraction of the modified
background are compared with the nominal values. 
Other contributions to the $R_{\rm em}$ systematic uncertainty
are negligible, when compared to the size of the total uncertainty.

\begin{figure}[H]
\begin{center}
\includegraphics[scale=0.40]{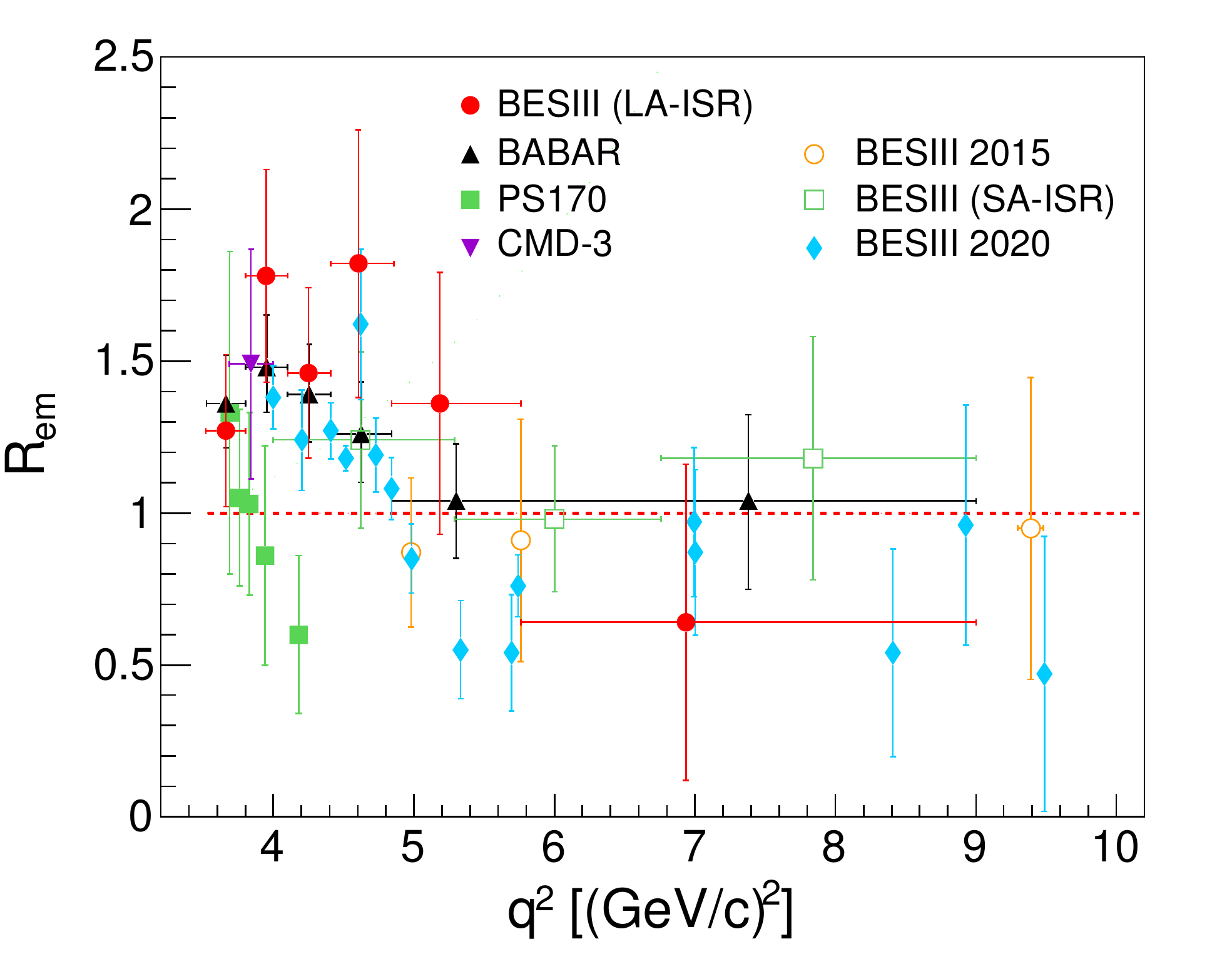}
\caption{Results for $R_{\rm em}$ from this work (BESIII (LA-ISR), red solid dots) 
as a function of the momentum transfer squared, $q^{2}$, together with the results 
from previous experiments: {\it BABAR}~\protect\cite{babar13t}, PS170~\protect\cite{Ba94}, CMD-3~\protect\cite{cmd3}, 
and BESIII~\protect\cite{bes3rscan,Ablikim:2019njl,Ablikim:2019eau}. 
Both statistical and systematic uncertainties are included
in all the results.}
 \label{fig:isrRatio}
\end{center}
\end{figure}

Figure~\ref{fig:isrRatio} shows the results for $R_{\rm em}$ 
from this analysis (red points), 
together with the results from previous measurements. 
Both statistical and systematic uncertainties are included. 
\section{Cross section for the process $e^{+}e^{-}\to p\bar{p}$ and proton effective form factor}
The Born cross section for the process $e^+ e^- \to p \bar p  $  is calculated in each $M_{p\bar p }$ interval $i$ and for each data sample $j$  ($j=1,2,...,7$) as follows:
\begin{linenomath*}
\begin{equation}
\sigma_{ij}=\frac{\mathcal{N}_{ij} - \mathcal{N}_{ij}^{\rm bkg} }{\epsilon_{ij} (1+\delta_{ij}) {\cal L}_{ij} },
\end{equation}
\end{linenomath*}
where  $\mathcal{N}_{ij}$ is the number of  selected $e^+ e^- \to p \bar p  \gamma$ candidates, $\mathcal{N}_{ij}^{\rm bkg}$ is the number of $e^{+}e^{-}\to p\bar{p}\pi^{0}$ background events, $\epsilon_{ij}$ is the detection efficiency,  $(1+\delta_{ij})$ is the radiative correction factor calculated from the MC simulations
and ${\cal L}_{ij}$ is the ISR binned integrated luminosity.  The  index $j$ runs over the seven c.m.  energies.  The binned integrated luminosity  ${\cal L}_{ij}$ is  calculated as:
\begin{linenomath*}
\begin{equation}
{\cal L}_{ij}=\int W(s_j, x_{ij}) {\cal {L}}_j dx_{ij},~ x_{ij}=1-\frac{q^2_{ij}}{s_j},
\label{diffli}
\end{equation}
\end{linenomath*}
where $W(s_j,x_{ij})$ [Eq.~(\ref{eq:isrxs})] is a function of the c.m. energy squared $s_j$ ($j=1,2,...,7$) and the energy fraction $x_{ij}$, and  ${\cal {L}}_j$  is the integrated luminosity collected at the c.m. energy $\sqrt{s_j}$ (Table~\ref{tab:isrData}).  
The integration in Eq.~(\ref{diffli}) 
is performed over the width of the selected $M_{p\bar p }$ interval.
The sum over the seven energy points for the binned integrated luminosity (${\cal L}_{i}$), the selected  $e^+ e^- \to p \bar p  \gamma$ candidates ($\mathcal{N}_{i}$) and the background events ($\mathcal{N}_{i}^{\rm bkg}$) in each $M_{p\bar p }$ interval is given in Table~\ref{tab:xsefRes}.  The averages over the seven c.m. energy points of the selection efficiencies ($\bar{\epsilon}_{i}$) and the radiative correction factors $\overline{(1+\delta_{i})}$ calculated as:
\begin{equation}
\begin{split}
\bar{\epsilon}_{i}&=\Sigma_j (\epsilon_{ij} {\cal {L}}_{ij})/{\cal {L}}_{i}, \\
\overline{(1+\delta_{i})}&=\Sigma_j((1+\delta_{ij}) {\cal {L}}_{ij})/{\cal {L}}_i,
\end{split}
\end{equation}
are also listed in Table~\ref{tab:xsefRes}. 
The Born cross sections $\sigma_{ij}$  are combined using 
the method of simple weighted averages~\cite{sch95}:
\begin{linenomath*}
\begin{equation} \label{eqxx}
\begin{split}
\sigma _{p \bar p }(M_{p\bar p })&=\sigma_i=   \Sigma_j (w_{ij} \sigma_{ij}),~\Delta \sigma_i=\sqrt{\frac{1}{\Sigma_j W_{ij}}}, \\
w_{ij}&=\frac{W_{ij}}{\Sigma_l W_{il}},~W_{ij}=\frac{1}{(\Delta \sigma_{ij})^2},
\end{split}
\end{equation}
\end{linenomath*}
where $\Delta \sigma$ is the statistical uncertainty of the cross section $\sigma$.   The  indices $j$ and $l$ run over the seven c.m.  energies. The systematic on the measurements of $\sigma_{ij}$ at the different c.m.energies are fully correlated and therefore need not be considered when determining the combination. The effective FF of the proton is calculated according to Eqs.~(\ref{eq:txs}) and (\ref{eq:geff}).

The experimental resolution on $M_{p\bar{p}}$ is typically eight times smaller
than the width of $M_{p\bar{p}}$ intervals and the event migration
across the intervals is relatively small, well below
the total uncertainty of the measured cross sections. The migration effect is taken
into account using the MC simulation.
\begin{figure*}[]
\begin{center}
\captionsetup[subfloat]{position=top,labelformat=empty}
  \subfloat[\label{fig:isrXS}]{\resizebox{0.5\textwidth}{!}{ \includegraphics{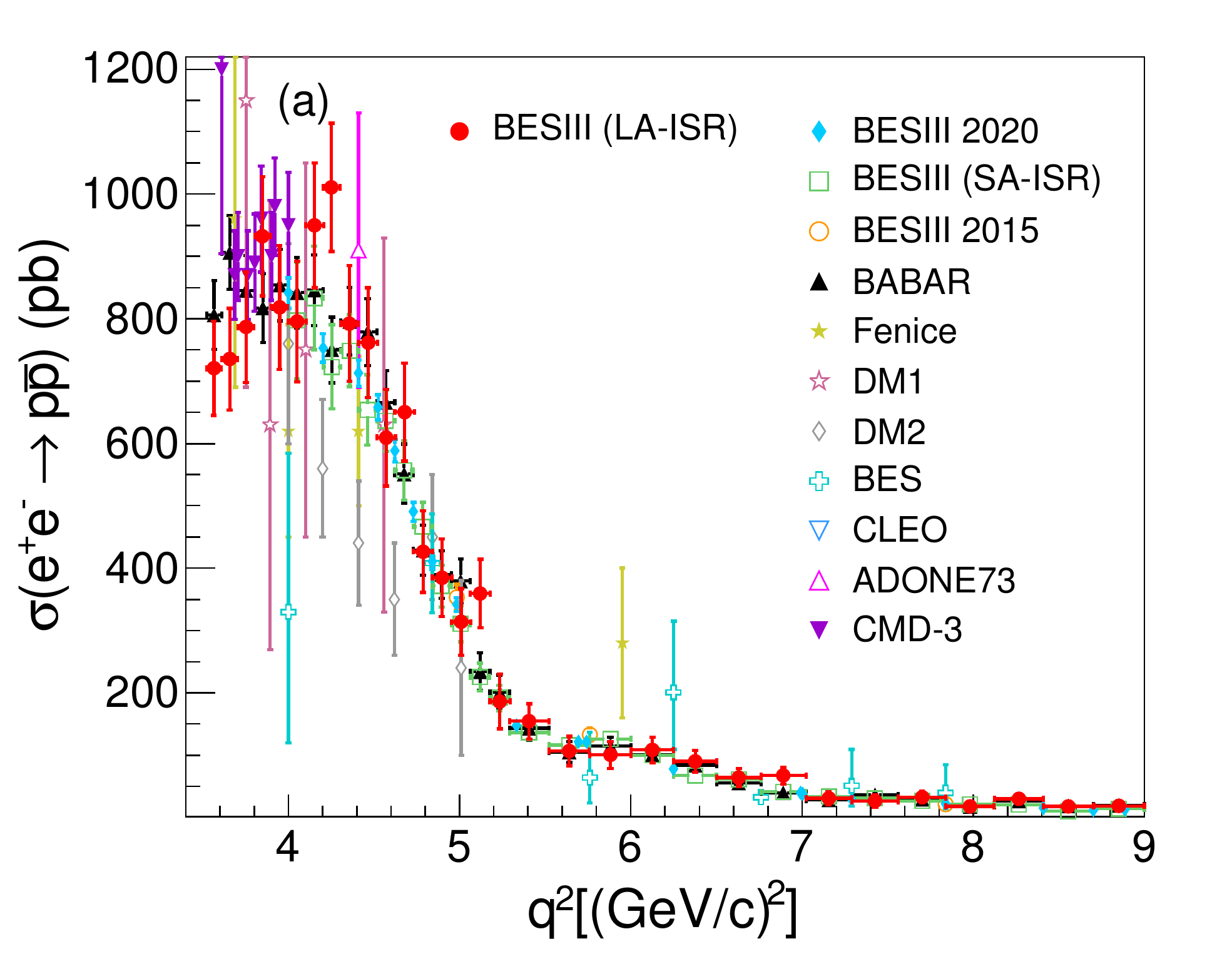} } }
  \subfloat[\label{fig:isrEff}]{\resizebox{0.5\textwidth}{!}{ \includegraphics{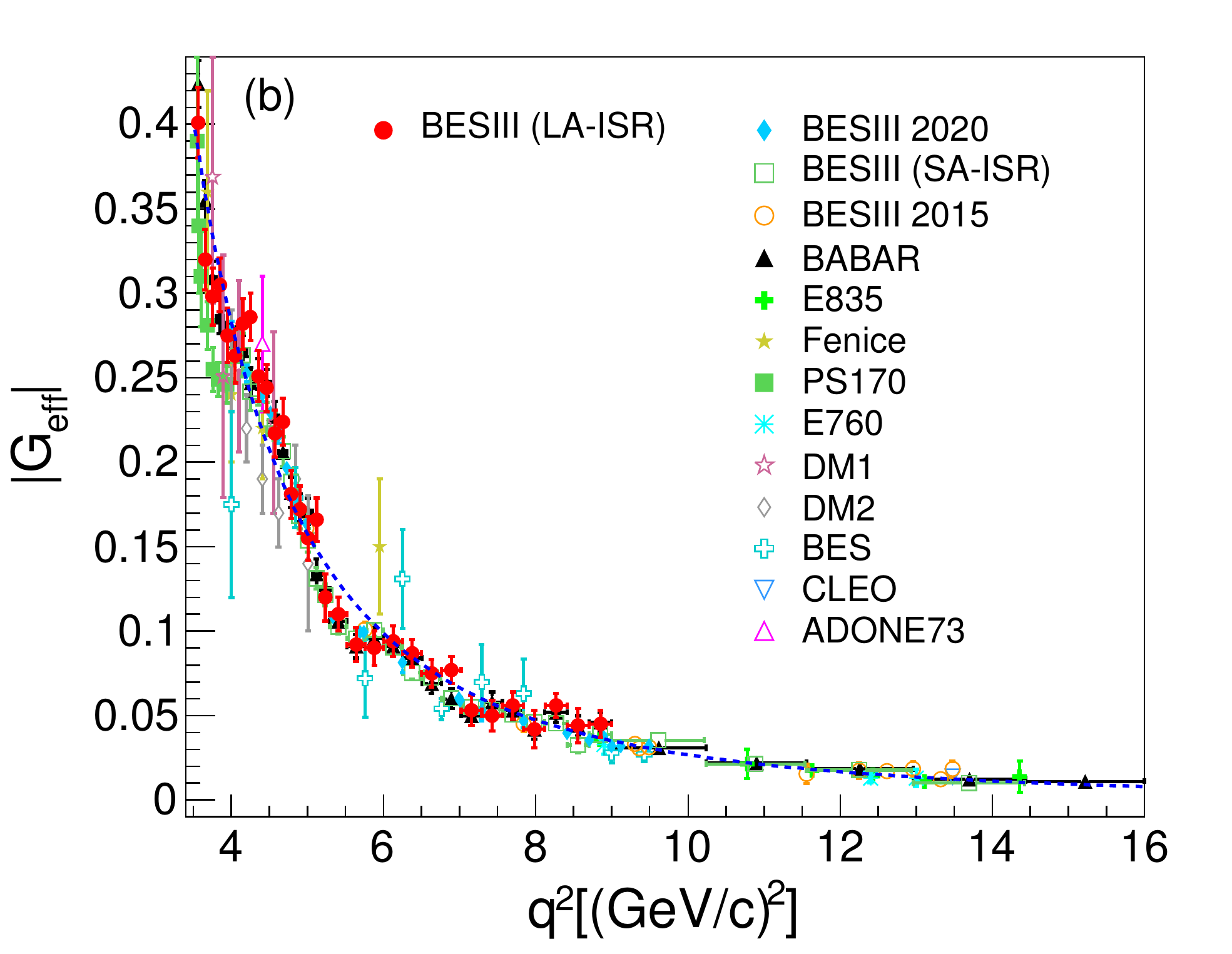} } }
  \caption{The cross section for the process $e^{+}e^{-}\to p\bar{p}$ (a) and the effective FF of the proton (b) measured in this work (red points), together with the results from previous experiments: BESIII~\protect\cite{bes3rscan,Ablikim:2019njl,Ablikim:2019eau}, {\it BABAR}~\protect\cite{babar13t,babar13u}, E835~\protect\cite{E835,Andreotti:2003bt}, Fenice~\protect\cite{fenice93, fenice94, fenice98}, PS170~\protect\cite{Ba94}, E760~\protect\cite{e760}, DM1~\protect\cite{dm182}, DM2~\protect\cite{dm283,dm290}, BES~\protect\cite{bes205},  CLEO~\protect\cite{cleo05}, and ADONE73~\protect\cite{adone73}. The  blue dashed curve shows the parameterization from Ref.~\cite{tomasi2001} based on Eq.~(\ref{rekalo}).}
\end{center}
\end{figure*}

Several sources are considered as contributing to the systematic uncertainties. 
The uncertainties from tracking and PID efficiencies and the $E/p$ requirement 
are each 1.0\% per track for all the $p\bar{p}$ mass intervals~\cite{bes3rscan}. 
The systematic uncertainty of the luminosity measurement is 1.0\% for all 
the data sets~\cite{ps2pLumi, xyzLumi}. The systematic uncertainties 
due to the 4C-kinematic fit and the background estimation are determined 
using the same methods as described in Section~\ref{ratioem}. 
The radiative correction factor $(1 + \delta)$ is calculated
in PHOKHARA generator with a theoretical uncertainty of 1\%.  The uncertainty from the energy dependence of the Born cross section used for the $(1 + \delta)$ calculation is determined by varying  the line shape of the cross section from PHOKHARA  event generator within the errors of the measured cross section.  The systematic uncertainties listed above are added in quadrature and are summarized in Table~\ref{tab:xsefSys}.

In Table~\ref{tab:xsefRes}, the obtained values of the Born cross section for the process $e^{+}e^{-}\to p\bar{p}$ and the effective FF of the proton are listed  including the statistical and systematic uncertainties. Figures~\ref{fig:isrXS} and \ref{fig:isrEff} show the results from this analysis for the $e^{+}e^{-}\to p\bar{p}$ cross section and the proton effective FF, respectively, together with results from previous experiments. 
\begin{table}[H]
\centering
\captionof{table}{Summary of the systematic uncertainty contributions (in $\%$)  
to the measurement of the Born cross section for the process $e^{+}e^{-}\to p\bar{p}$. 
The contributions from the tracking and PID efficiencies, $E/p$ requirement,
radiative corrections 
and luminosity are uniform over the considered $p\bar{p}$ mass range.  
The systematic uncertainty due to the 4C kinematic fit and background subtraction 
depends on the $p\bar{p}$ mass interval. The systematic uncertainties are added in quadrature (last raw of the table).
}
\begin{tabular*}{0.475\textwidth}{l @{\extracolsep{\fill}} r}
 \toprule
 \hline
  \multicolumn{1}{c}{Source}  &  \multicolumn{1}{c}{Uncertainty  (\%)}   \\
 \midrule
  Tracking efficiency  &  2 \\
  PID efficiency       &  2 \\
  $E/p$                &  1 \\
  Luminosity           &  1 \\
  Radiative corrections & 1 - 4 \\
  4C kinematic fit     &  1 - 4 \\
  Background subtraction & 1 - 8 \\
  \hline
  Sum (cross section)   &  4 - 9  \\
 \hline
 \bottomrule
\end{tabular*}
\label{tab:xsefSys}
\end{table}
\begin{figure}[H]
\begin{center}
  \includegraphics[scale=0.40]{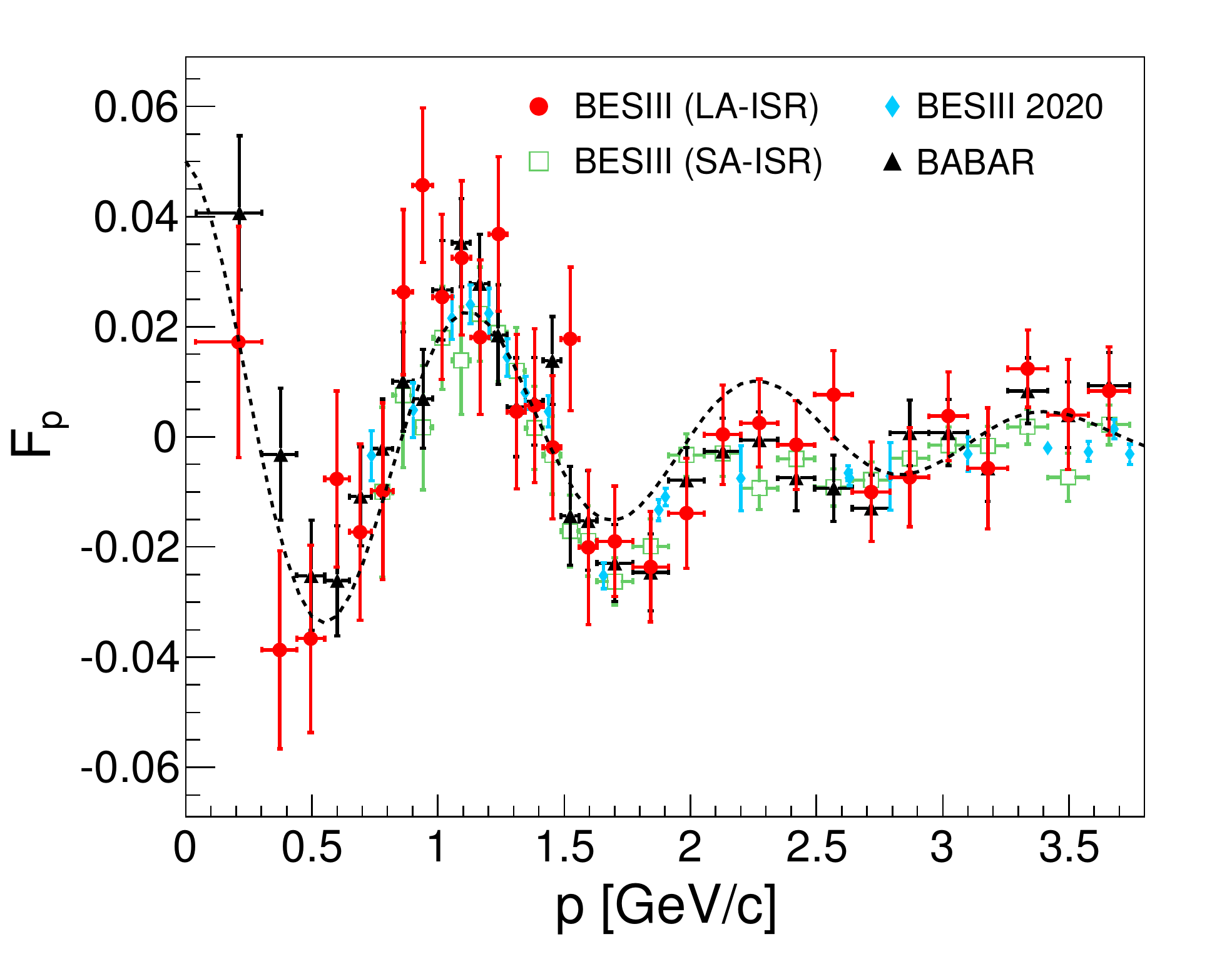}
 \caption{The effective FF of the proton, after subtraction of the smooth function described by Eq.~(\ref{rekalo}), as a function of the relative momentum ${p}$.  The data are from the present analysis (red points) and previous measurements of BESIII~\protect\cite{Ablikim:2019njl,Ablikim:2019eau} and {\it BABAR}~\protect\cite{babar13t,babar13u}. The  black dashed curve shows the parameterization from Ref.~\cite{Bianconi2016} based on Eq.~(\ref{fosci}).}
 \label{GEFFTOTPIDOSC}
\end{center}
\end{figure}

The data on the TL effective FF are best reproduced by the function proposed in Ref.~\cite{tomasi2001},
\begin{linenomath*}
\begin{equation}
|G_{\rm eff}|=\frac{{\cal{A}}}{(1+q^2/m_a^2)[1-q^2/q_0^2 ]^2},~q_0^2=0.71~(\mbox{GeV/$c$})^2,
\label{rekalo}
\end{equation}
\end{linenomath*}
where ${\cal{A}}=7.7$ and $m_a^2=14.8~(\mbox{GeV/$c$})^2$ are the fit parameters obtained previously in Ref.~\cite{Bianconi2015}. This function is illustrated in Fig.~\ref{fig:isrEff} by the blue dashed curve and  reproduces the behavior of the effective FF over the full $q^2$ range. However, the measurements indicate oscillating structures which are clearly seen when the residuals are plotted as a function of the  relative momentum ${p}$ of the final proton and antiproton \cite{Bianconi2016}. Figure~\ref{GEFFTOTPIDOSC} shows the values of the proton effective FF as a function of ${p}$  after subtraction of the smooth function described by Eq.~(\ref{rekalo}). The black curve in Fig.~\ref{GEFFTOTPIDOSC} describes the periodic oscillations and has the form \cite{Bianconi2016}
\begin{linenomath*}
\begin{equation}
F_{p}=A^{\rm osc} \exp(-B^{\rm osc} {p}) \cos (C^{\rm osc} {p} +D^{\rm osc}),
\label{fosci}
\end{equation}
\end{linenomath*}
where $A^{\rm osc}=0.05$, $B^{\rm osc}=0.7~(\mbox{GeV/$c$})^{-1}$, $C^{\rm osc}=5.5~(\mbox{GeV/$c$})^{-1}$ and $D^{\rm osc}=0.0$ have been obtained from a fit to the {\it BABAR} data \cite{Bianconi2015}. 

\section{Summary}
Using seven data sets with a total integrated luminosity 
of 7.5~fb$^{-1}$ collected by the BESIII experiment 
at $\sqrt{s}$ between 3.773 and 4.600 GeV, 
the ratio of the proton electromagnetic FF absolute values, 
the Born cross section for the process  $e^+e^-\to p\bar{p}$ 
and the effective FF of the proton are measured from the $p\bar{p}$ threshold 
to 3.0 GeV/$c^{2}$ through the ISR process $e^+e^-\to p\bar{p}\gamma$. 
This measurement confirms an enhancement 
of the ratio of FFs in the $M_{p\bar{p}}$
region below 2.2 GeV/$c^{2}$ previously observed by {\it BABAR} and BESIII
and differs from the behavior reported by PS170 \cite{Ba94}.
Close to the threshold, the observed ratio is compatible with unity within the uncertainties.
The results on the Born cross section 
for the process $e^{+}e^{-}\to p\bar{p}$ and the proton effective FF presented in this work 
are in a good agreement 
with the measurements from the previous 
experiments~\cite{adone73, dm182, dm283, dm290, fenice93, fenice94, fenice98, cleo05, bes205, babar06, babar13t, bes3rscan, Ablikim:2019njl, Ablikim:2019eau, cmd3, e760, Ba94, E835,Andreotti:2003bt}. 
In particular, we reproduce the structures seen in the {\it BABAR} and previous BESIII 
measurements of the proton effective FF. The origin of these oscillating structures 
can be attributed to an interference effect involving rescattering processes in the 
final state \cite{Bianconi2015} or to independent resonant structures, 
as in Ref.~\cite{Lorentz2015}. The precision of the measurements obtained 
in this work are comparable to or lower than that achieved in previous 
BESIII studies~\cite{bes3rscan, Ablikim:2019njl, Ablikim:2019eau} using 
the direct annihilation and SA-ISR processes which benefit from higher statistics. 
The analysis described here shows the possibility to use the LA-ISR  technique at BESIII 
to perform independent and complementary measurements of the proton FFs 
down to the production threshold. 
Larger sampes that are  currently being collected by BESIII~\cite{bes2020} 
will enhance the precision of these measurements.
\begin{table*}
\centering
\small
 \captionof{table}{The integrated luminosity (${\cal L}_{i}$), the number of candidates ( $\mathcal{N}_{i}$), the estimated $e^{+}e^{-}\to p\bar{p}\pi^{0}$ background ($\mathcal{N}_{i}^{\rm bkg}$), the average over the seven c.m. energy points of the selection efficiencies ($\bar{\epsilon}_{i}$) and the radiative correction factors ($\overline{(1+\delta_{i})}$), the measured $e^+e^-\to p\bar{p}$ Born cross section ($\sigma_{i}$) and the effective FF ($|G_{\rm eff}|$).}
  \begin{tabular}{cccccccc}
  \toprule
  \hline
  $ M_{p\bar{p}}$~[GeV/$c^{2}$]   &  ${\cal L}_{i}$~[pb$^{-1}$]  &  $\mathcal{N}_{i}$ &  $\mathcal{N}_{i}^{\rm bkg}$ &  $\bar{\epsilon}_{i}$ &  $\overline{(1+\delta_{i})}$  & $\sigma_{i}$~[pb]  & $|G_{\rm eff}|$  
%
 %
\\
  \midrule
1.877 - 1.900 &  2.11 &   167 &   30.8 $\pm$   2.3 & 0.071 &  1.21 &   721 $\pm$    71 $\pm$    26 & 0.401 $\pm$ 0.020 $\pm$ 0.007 \\
1.900 - 1.925 &  2.30 &   194 &   52.8 $\pm$   3.0 & 0.068 &  1.15 &   735 $\pm$    77 $\pm$    26 & 0.320 $\pm$ 0.017 $\pm$ 0.006 \\
1.925 - 1.950 &  2.36 &   223 &   77.4 $\pm$   3.6 & 0.069 &  1.13 &   787 $\pm$    84 $\pm$    30 & 0.298 $\pm$ 0.016 $\pm$ 0.006 \\
1.950 - 1.975 &  2.41 &   263 &   88.8 $\pm$   3.9 & 0.068 &  1.12 &   933 $\pm$    90 $\pm$    34 & 0.305 $\pm$ 0.015 $\pm$ 0.006 \\
1.975 - 2.000 &  2.47 &   270 &  113.0 $\pm$   4.4 & 0.067 &  1.12 &   818 $\pm$    90 $\pm$    41 & 0.275 $\pm$ 0.015 $\pm$ 0.007 \\
2.000 - 2.025 &  2.53 &   269 &  114.6 $\pm$   4.4 & 0.067 &  1.11 &   796 $\pm$    89 $\pm$    38 & 0.263 $\pm$ 0.015 $\pm$ 0.006 \\
2.025 - 2.050 &  2.59 &   309 &  118.9 $\pm$   4.5 & 0.066 &  1.11 &   950 $\pm$    93 $\pm$    38 & 0.282 $\pm$ 0.014 $\pm$ 0.006 \\
2.050 - 2.075 &  2.65 &   321 &  123.1 $\pm$   4.6 & 0.066 &  1.11 &  1011 $\pm$    95 $\pm$    41 & 0.286 $\pm$ 0.013 $\pm$ 0.006 \\
2.075 - 2.100 &  2.72 &   282 &  112.8 $\pm$   4.4 & 0.065 &  1.11 &   792 $\pm$    86 $\pm$    36 & 0.251 $\pm$ 0.014 $\pm$ 0.006 \\
2.100 - 2.125 &  2.79 &   264 &  105.7 $\pm$   4.2 & 0.066 &  1.10 &   762 $\pm$    82 $\pm$    31 & 0.244 $\pm$ 0.013 $\pm$ 0.005 \\
2.125 - 2.150 &  2.85 &   227 &   92.7 $\pm$   3.9 & 0.066 &  1.10 &   610 $\pm$    73 $\pm$    25 & 0.217 $\pm$ 0.013 $\pm$ 0.004 \\
2.150 - 2.175 &  2.93 &   236 &   92.4 $\pm$   3.9 & 0.065 &  1.10 &   650 $\pm$    74 $\pm$    26 & 0.224 $\pm$ 0.013 $\pm$ 0.004 \\
2.175 - 2.200 &  3.00 &   178 &   81.2 $\pm$   3.7 & 0.066 &  1.10 &   427 $\pm$    62 $\pm$    19 & 0.181 $\pm$ 0.013 $\pm$ 0.004 \\
2.200 - 2.225 &  3.08 &   174 &   83.7 $\pm$   3.7 & 0.066 &  1.10 &   385 $\pm$    60 $\pm$    17 & 0.172 $\pm$ 0.013 $\pm$ 0.004 \\
2.225 - 2.250 &  3.15 &   140 &   65.2 $\pm$   3.3 & 0.067 &  1.10 &   314 $\pm$    52 $\pm$    14 & 0.155 $\pm$ 0.013 $\pm$ 0.003 \\
2.250 - 2.275 &  3.24 &   152 &   65.4 $\pm$   3.3 & 0.068 &  1.10 &   359 $\pm$    53 $\pm$    15 & 0.166 $\pm$ 0.012 $\pm$ 0.004 \\
2.275 - 2.300 &  3.32 &   114 &   62.1 $\pm$   3.2 & 0.068 &  1.09 &   186 $\pm$    43 $\pm$    11 & 0.120 $\pm$ 0.014 $\pm$ 0.003 \\
2.300 - 2.350 &  6.91 &   192 &  105.6 $\pm$   4.2 & 0.070 &  1.10 &   154 $\pm$    27 $\pm$     8 & 0.110 $\pm$ 0.010 $\pm$ 0.003 \\
2.350 - 2.400 &  7.28 &   157 &   93.5 $\pm$   3.9 & 0.072 &  1.09 &   107 $\pm$    23 $\pm$     6 & 0.092 $\pm$ 0.010 $\pm$ 0.003 \\
2.400 - 2.450 &  7.69 &   149 &   82.1 $\pm$   3.7 & 0.073 &  1.09 &   100 $\pm$    20 $\pm$     8 & 0.090 $\pm$ 0.009 $\pm$ 0.003 \\
2.450 - 2.500 &  8.13 &   139 &   66.9 $\pm$   3.3 & 0.075 &  1.08 &   108 $\pm$    19 $\pm$     7 & 0.094 $\pm$ 0.008 $\pm$ 0.003 \\
2.500 - 2.550 &  8.60 &   126 &   62.8 $\pm$   3.2 & 0.076 &  1.09 &    90 $\pm$    17 $\pm$     5 & 0.087 $\pm$ 0.008 $\pm$ 0.002 \\
2.550 - 2.600 &  9.12 &   104 &   54.0 $\pm$   3.0 & 0.077 &  1.08 &    64 $\pm$    14 $\pm$     3 & 0.075 $\pm$ 0.008 $\pm$ 0.002 \\
2.600 - 2.650 &  9.68 &   109 &   52.9 $\pm$   2.9 & 0.077 &  1.09 &    67 $\pm$    13 $\pm$     3 & 0.077 $\pm$ 0.008 $\pm$ 0.002 \\
2.650 - 2.700 & 10.29 &    75 &   47.3 $\pm$   2.8 & 0.078 &  1.08 &    31 $\pm$    11 $\pm$     2 & 0.053 $\pm$ 0.009 $\pm$ 0.001 \\
2.700 - 2.750 & 10.97 &    73 &   47.4 $\pm$   2.8 & 0.080 &  1.07 &    26 $\pm$    10 $\pm$     2 & 0.050 $\pm$ 0.009 $\pm$ 0.002 \\
2.750 - 2.800 & 11.71 &    82 &   48.4 $\pm$   2.8 & 0.080 &  1.07 &    33 $\pm$    10 $\pm$     2 & 0.056 $\pm$ 0.008 $\pm$ 0.002 \\
2.800 - 2.850 & 12.54 &    70 &   50.6 $\pm$   2.8 & 0.080 &  1.08 &    18 $\pm$     9 $\pm$     1 & 0.042 $\pm$ 0.011 $\pm$ 0.002 \\
2.850 - 2.900 & 13.45 &    92 &   47.5 $\pm$   2.7 & 0.080 &  1.06 &    30 $\pm$     8 $\pm$     3 & 0.056 $\pm$ 0.007 $\pm$ 0.002 \\
2.900 - 2.950 & 14.47 &    69 &   49.9 $\pm$   2.8 & 0.081 &  1.07 &    18 $\pm$     8 $\pm$     1 & 0.044 $\pm$ 0.010 $\pm$ 0.002 \\
2.950 - 3.000 & 15.62 &    71 &   49.3 $\pm$   2.8 & 0.082 &  1.06 &    19 $\pm$     7 $\pm$     1 & 0.045 $\pm$ 0.008 $\pm$ 0.001 \\
  \hline
  \bottomrule
 \end{tabular}
 \label{tab:xsefRes}
\end{table*}
\normalsize
%
\section*{Acknowledgement}
The BESIII collaboration thanks the staff of BEPCII and the IHEP computing center for their strong support. This work is supported in part by National Natural Science Foundation of China (NSFC) under Contracts Nos. 11625523, 11635010, 11735014, 11822506, 11835012, 11935015, 11935016, 11935018, 11961141012; the Chinese Academy of Sciences (CAS) Large-Scale Scientific Facility Program; Joint Large-Scale Scientific Facility Funds of the NSFC and CAS under Contracts Nos. U1732263, U1832207; CAS Key Research Program of Frontier Sciences under Contracts Nos. QYZDJ-SSW-SLH003, QYZDJ-SSW-SLH040; 100 Talents Program of CAS; INPAC and Shanghai Key Laboratory for Particle Physics and Cosmology; ERC under Contract No. 758462; German Research Foundation DFG under Contracts Nos. 443159800, Collaborative Research Center CRC 1044, FOR 2359, FOR 2359, GRK 214; Istituto Nazionale di Fisica Nucleare, Italy; Ministry of Development of Turkey under Contract No. DPT2006K-120470; National Science and Technology fund; Olle Engkvist Foundation under Contract No. 200-0605; STFC (United Kingdom); The Knut and Alice Wallenberg Foundation (Sweden) under Contract No. 2016.0157; The Royal Society, UK under Contracts Nos. DH140054, DH160214; The Swedish Research Council; U. S. Department of Energy under Contracts Nos. DE-FG02-05ER41374, DE-SC-0012069.
\bibliography{mybibfile}
\bibliographystyle{elsarticle-num}
%
\end{multicols}

\end{document}